 \documentclass[final,5p,times,twocolumn]{elsarticle}
\usepackage{amsmath,amssymb}
\journal{NIMA proceedings}
\usepackage{graphicx}
\usepackage{color}
\usepackage[section]{placeins}

\begin{document}

\begin{frontmatter}

\title{Tunable High Gradient Quadrupoles For A Laser Plasma Acceleration Based FEL\tnoteref{mytitlenote}}

\author{A. Ghaith\fnref{myfootnote}}
\author{C. Kitegi\fnref{myfootnote}}
\author{T. Andr\'e\fnref{myfootnote}}
\author{M. Vall\'eau\fnref{myfootnote}}
\author{F. Marteau\fnref{myfootnote}}
\author{J. V\'et\'eran\fnref{myfootnote}}
\author{F. Blache\fnref{myfootnote}}
\author{C. Benabderrahmane\fnref{myfootnote2}}
\author{O. Cosson\fnref{myfootnote3}}
\author{F. Forest\fnref{myfootnote3}}
\author{P. Jivkov\fnref{myfootnote3}}
\author{J. L. Lancelot\fnref{myfootnote3}}
\author{M. E. Couprie\fnref{myfootnote}}

\address{L'Orme des Merisiers, 91190 Saint-Aubin}
\fntext[myfootnote]{Synchrotron SOLEIL}
\fntext[myfootnote2]{European Synchrotron Radiation Facility}
\fntext[myfootnote3]{SIGMAPHI}

\begin{abstract}
Laser Plasma Acceleration (LPA) is capable of producing a GeV beam within a cm accelerating distance, but with a rather high initial divergence and large energy spread. COXINEL aims to demonstrate a compact Free Electron Laser using such a source, where a specific transport line with adequate elements is used, such as tunable high gradient quadrupoles for handling the divergence. An innovative permanent magnet based quadrupole (QUAPEVA) made of two quadrupoles superimposed capable of generating a gradient of 200 T/m is presented. The first quadrupole consists of magnets shaped as a ring and attaining a constant gradient of 155 T/m, and the second one made of four cylindrical magnets surrounding the ring and capable of rotating around their axis to achieve a gradient tunability of $\pm$ 46 T/m. Each tuning magnet is connected to a motor and controlled independently, enabling the gradient to be tuned with a rather good magnetic center stability ($\pm$10 $\mu$m) and without any field asymmetry. The measurements and field optimization of seven quadrupoles with different magnetic lengths are reported. A set of QUAPEVA triplet, installed at COXINEL, achieved good focusing and enabled beam based alignment.
\end{abstract}

\begin{keyword}
\texttt{Permanent magnet, quadrupole}
\MSC[2010] 00-01\sep  99-00
\end{keyword}

\end{frontmatter}

\section{Introduction}

Laser Plasma Acceleration (LPA) performance has exceeded that of conventional RF accelerators regarding accelerating distance\cite{balakin1995focusing, OidePhysRevAccelBeams.19.111005}. It can now generate several GeV beam within a very short accelerating distance (few centimeters), high peak current $\sim$10 kA and short bunch duration (few fs), however the divergence is quite large (few mrads) as well as the energy spread (few percent). Thus this application has a requirement of quadrupoles with gradient up to 200 T/m to handle such a beam. With the use of permanent magnet technology, one is able to reduce the bore aperture of the quadrupole and thus attain a larger gradient. Permanent Magnet Quadrupoles (PMQs) achieve gradients of the order of hundreds T/m with compactness, and also with the absence of power supplies makes them a solution for future sustainable green society.

Several Halbach \cite{halbach1983permanent} ring based PMQs with fixed gradient were designed and built: at CESR  \cite{lou1998stability}; at Kyoto University / SLAC  \cite{mihara2004super}; at CORNELL  \cite{lim2005adjustable}; at the department f$\ddot{u}$r Physik \cite{eichner2007miniature}; at ESRF  \cite{Ngotta2016hybrid}. As for introducing gradient tunability in PMQs, original designs were proposed and developed to provide a variable gradient, such as at SLAC / Fermilab collaboration \cite{gottschalk2005performance}; at Kyoto U. / SLAC collaboration \cite{mihara2006variable}; at STFC Daresbury Laboratory / CERN for the CLIC project  \cite{shepherd2012novel}.

In this paper, a hybrid permanent magnet based quadrupole with tunable gradient (QUAPEVA) developed at Synchrotron SOLEIL in collaboration with SIGMAPHI is presented. Its design provides a high tunable gradient strength 156 $\pm$ 46 T/m and small magnetic center excursions within $\pm$10 $\mu m$ in both transverse planes. Seven systems have been built with different magnetic lengths, a first set of triplet (26 mm, 40.7 mm, 44.7 mm), a second triplet (47.1 mm, 61 mm, 81.1 mm), and a prototype (100 mm). Simulation models are presented alongside the mechanical design. Different magnetic measurements are used to characterize the field quality and are compared with the simulation models. Finally, three QUAPEVAs have been used for the COXINEL project \cite{couprie2014towards,couprie2016application}.

\section{\label{sec:level1}Magnetic Design}

\begin{figure}[h!]
\centering
\includegraphics[scale=0.5]{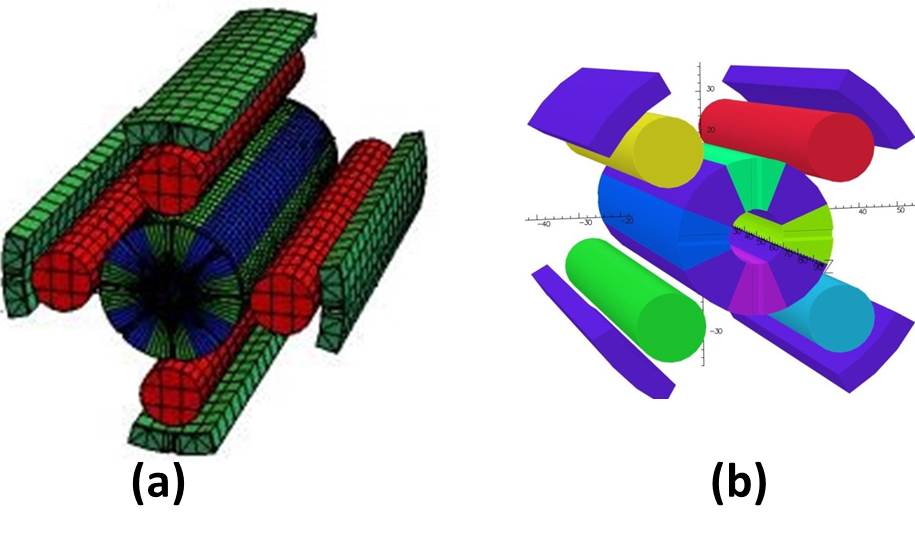}
\caption{(a) RADIA model, (b) TOSCA model.}
\label{Fig1}
\end{figure}

The QUAPEVA consists of two superimposed quadrupoles, where the first quadrupole placed in the center is a Halbach hybrid ring with a fixed gradient surrounded by a second one composed of four cylindrical magnets capable of rotating around their axis providing the gradient tunability. The geometry and magnetic characteristics of the QUAPEVA are optimized using two numerical tools: RADIA \cite{RADIAChubarJSR1998} a magnetostatic  code based on boundary integral method(see Fig. \ref{Fig1}-a) and TOSCA \cite{TOSCA} a finite element magnetostatic code (see Fig. \ref{Fig1}-b). The magnet and pole characteristics are shown in Table \ref{T1}.

\begin{table}[h]
	\small
	\centering
	\caption{QUAPEVA characteristics.}
		\begin{tabular}[c]{ccc}
		\hline
		\textbf{Parameter}&\textbf{Value}&\textbf{Unit}\\
		\hline
Remanent Field& 1.26 & T\\
Coercivity& 1830 & kA/m\\
Pole saturation &2.35& T\\
 Radius for Good Field Region (GFR)&4&mm\\
$\Delta G/G$ at GFR&$< 0.01$& \\[1ex]
 \hline
		\hline
		\end{tabular}
	\label{T1}
\end{table}


\begin{figure}[h!]
\centering
\includegraphics[scale=0.6]{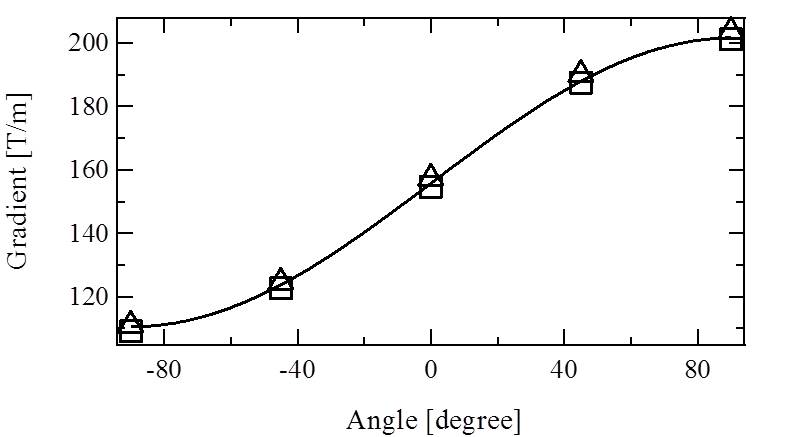}
\caption{Gradient evolution computed for the prototype QUAPEVA versus tuning magnets angle from -90$^o$ to 90$^o$. ($\triangle$) TOSCA and ($\square$) RADIA. (Line) sinus fit.}
\label{Fig2}
\end{figure}

\FloatBarrier
Fig. \ref{Fig2} shows the simulated gradient evolution with tuning magnet angle in the case of the prototype, where the gradient reaches a maximum and a minimum value between -90 and +90 degrees. Simulation results of RADIA and TOSCA models are in good agreement. The evolution is fitted with a sinus function $G(\theta)=G_0 + G_t sin(\theta)$, where $G_0$ is the gradient contribution of the main magnets (tuning magnets are in their reference position), $G_t$ the gradient contribution of all the tuning magnets, and $\theta$ their corresponding angle. In the case of the prototype (100 mm magnetic length), $G_0$=155 T/m and $G_t$=46 T/m are found. The gradient variation from peak to peak is $\sim$92 T/m and the maximum gradient reached is $\sim$201 T/m. The maximum gradient and tunability of each system are shown in Table. \ref{T2}.

\begin{table}[h]
	\small
	\centering
	\caption{Maximum gradient and tunability for the systems built.}
		\begin{tabular}[c]{ccc}
		\hline
		\textbf{Magnetic length}&\textbf{$G_{max}$ [T/m]}&\textbf{$\Delta G$ [T/m]}\\
		\hline
100 mm& 201 & 92\\ [0.5ex] 
81.1 mm& 195 & 89 \\  
61 mm& 190&88 \\ 
47.1 mm&184 &86\\
44.7 mm&183&86\\
40.7 mm&180&85\\
26 mm&164&78\\[1ex]
 \hline
		\hline
		\end{tabular}
	\label{T2}
\end{table}

\section{\label{sec:level1}Mechanical Design}
\begin{figure}[h!]
\centering
\includegraphics[scale=0.5]{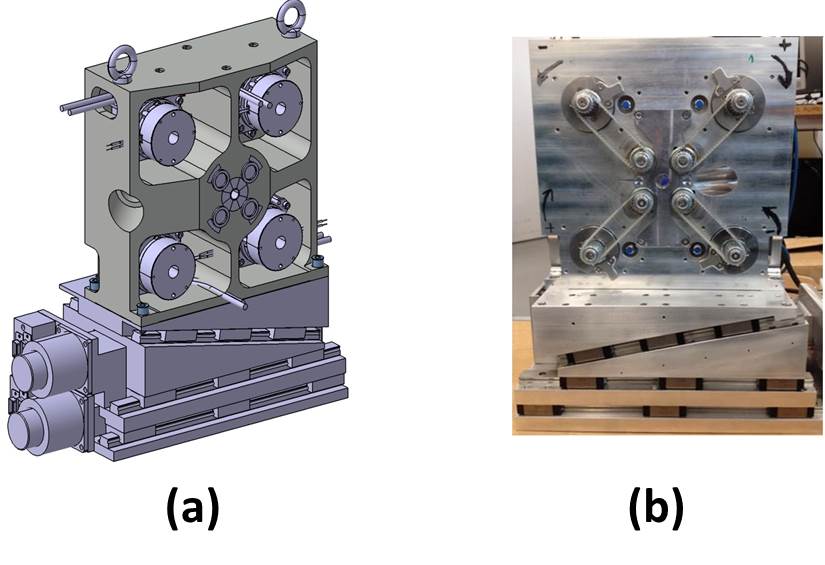}
\caption{a) mechanical design of the QUAPEVA, (b) an assembled QUAPEVA.}
\label{Fig3}
\end{figure}

The quadrupole is supported by a translation table to compensate for the residual shift of the magnetic center in both transverse planes as the gradient is varied, and is built into an Aluminum support frame to maintain the magnetic elements in their positions. Each cylindrical magnet is connected to one motor preventing a break of symmetry whilst changing the gradient. The motors are placed at the corners to avoid perturbations of the magnetic field, and are connected to the cylindrical magnets by non-magnetic belts that transmit the rotation movement. The chosen motors (HARMONIC DRIVE, FHA-C mini motors) have sufficient torque to counteract the magnetic forces, are very compact (48.5 x 50 x 50 mm$^3$), and have an encoder of 31 $\mu rad$ resolution. Fig. \ref{Fig3} presents the mechanical design (a) along side an assembled QUAPEVA (b).

\section{\label{sec:level1} Magnetic measurements of QUAPEVA}

\begin{figure}[ht]
\centering
\includegraphics[scale=0.05]{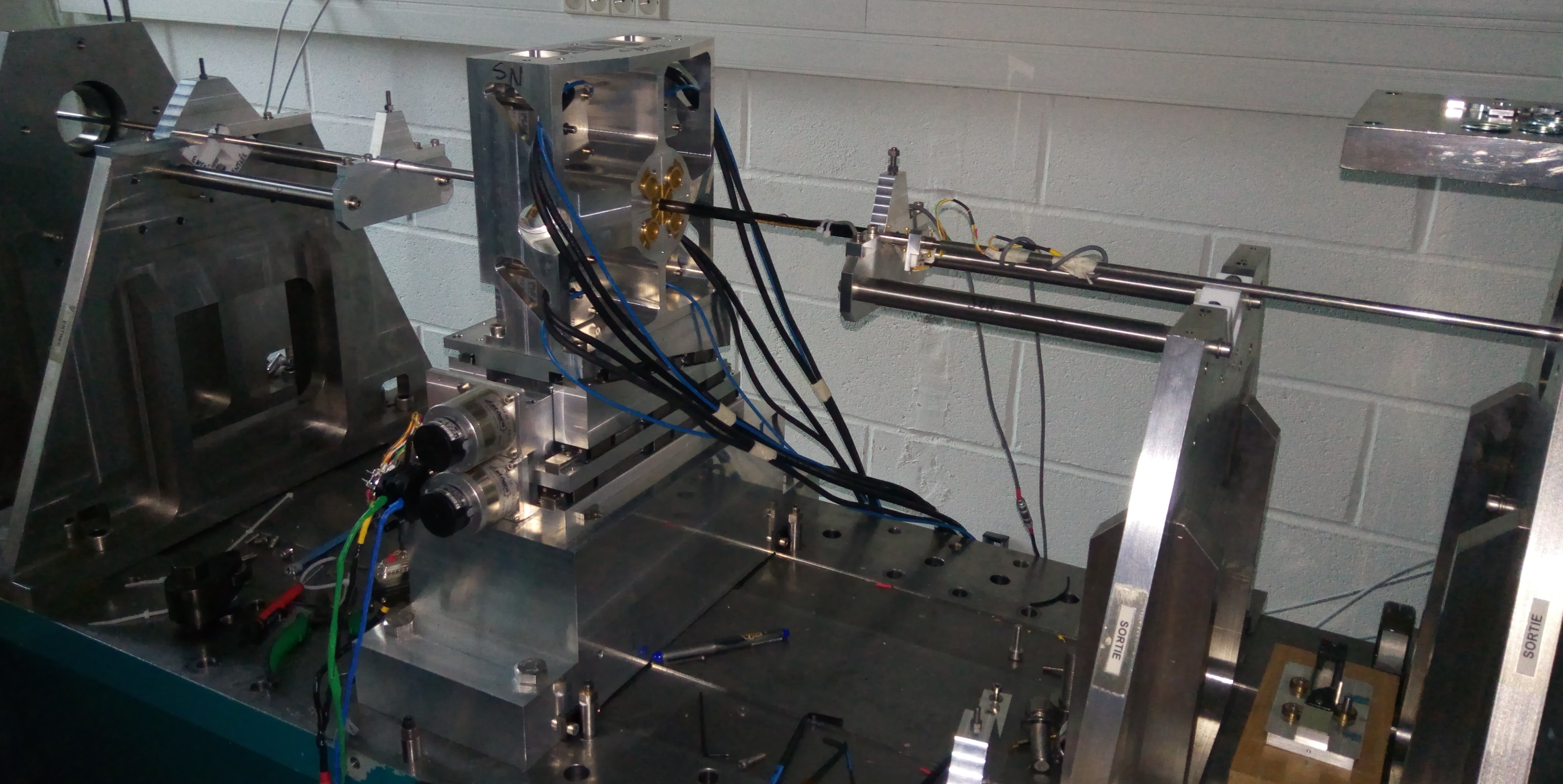}
\caption{\label{Fig4}QUAPEVA installed on SOLEIL rotating coil bench.}
\end{figure}

Different measurement methods have been used to characterize the magnetic field of the QUAPEVA. A rotating coil, built for the SOLEIL magnet characterization bench \cite{madur2006contribution} shown in Fig. \ref{Fig4}, to measure the field integral, and thus determine the gradient and magnetic center. A single stretched wire to measure the field integral which is similar to the rotating coil method concept. A pulsed wire method, where one applies a square current pulse through a wire and the use of a laser sensor to track the wire deflection that is proportional to the field\cite{preston1992wiggler}.

\FloatBarrier

\begin{figure}[ht]
\centering
\includegraphics[scale=0.5]{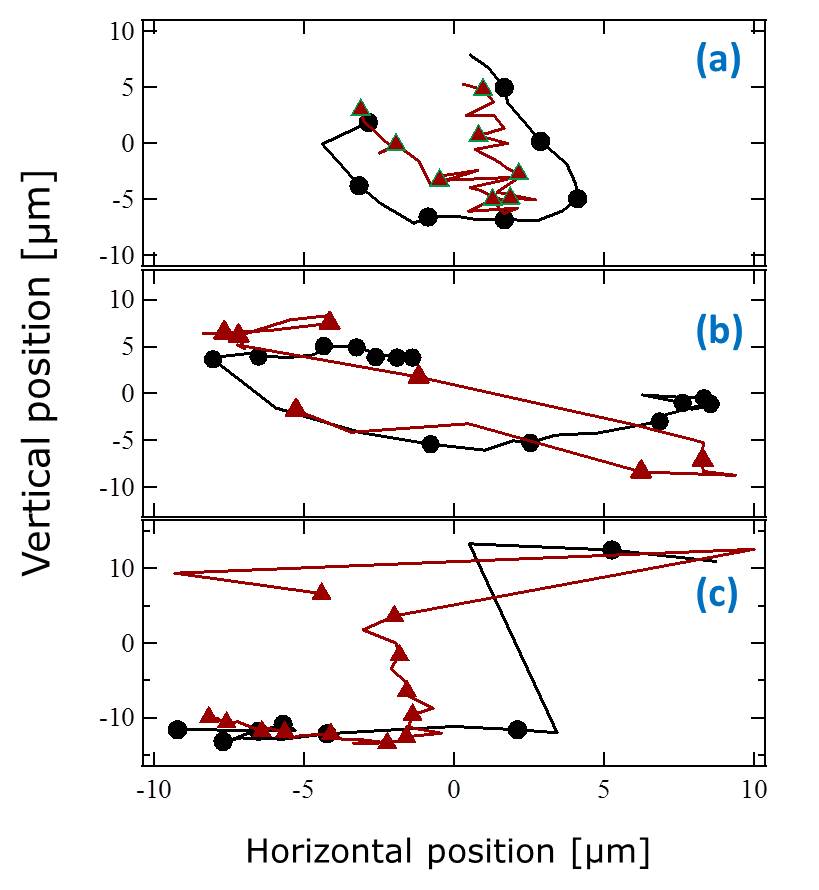}
\caption{\label{Fig5}(a) 26 mm, (b) 40.7 mm, (c) 44.7 mm.}
\end{figure}

While varying the gradient, some asymmetry will occur and cause magnetic center excursion. This excursion should be kept as small as possible due to the fact that the QUAPEVAs are used to focus a large energy spread beam, and if one tunes the gradient to focus a different energy the magnetic center should not change much. Fig. \ref{Fig5} shows the magnetic center excursion measurements for the first triplet, as the gradient is varied from minimum to maximum, using two methods stretched wire ({\color{red}$\blacktriangle$}) and rotating coil ($\bullet$). The two measurements are in good agreement where the magnetic center excursion is kept within $\pm$10 $\mu$m in both axes.

\begin{figure}[ht]
\centering
\includegraphics[scale=0.6]{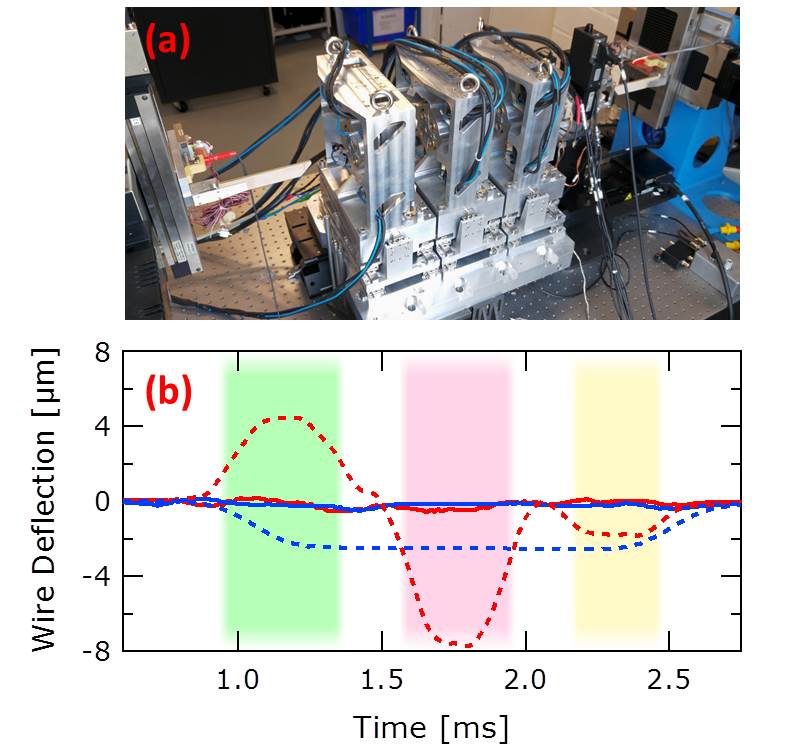}
\caption{(a) Triplet of QUAPEVAs installed on the pulsed wire bench for magnet center alignment. (b) wire displacement through a triplet set of QUAPEVAs. (Dashed) before the alignment, (line) after alignment, (red) vertical axis, (Blue) horizontal.}
\label{Fig6}
\end{figure}

 Before the commissioning of the first triplet, one must assure that the magnetic centers of the three QUAPEVAs are well aligned. Fig. \ref{Fig6}-a shows the triplet installed at the pulsed wire measurement bench, to measure the magnetic centers of the QUAPEVAs. The QUAPEVAs were centered one by one, starting from the 40.7 mm, the 44.7 mm and finally the 26 mm. Fig. \ref{Fig6}-b shows the wire deflection versus time before (dashed) and after (line) the adjustments of the translation tables. As the field is proportional to the displacement, the magnetic centers are well aligned when the deflection is $\sim$ 0. Thus good alignment has been achieved.

\FloatBarrier
\section{Application to COXINEL }

\begin{figure}[ht]
\centering
\includegraphics[scale=0.25]{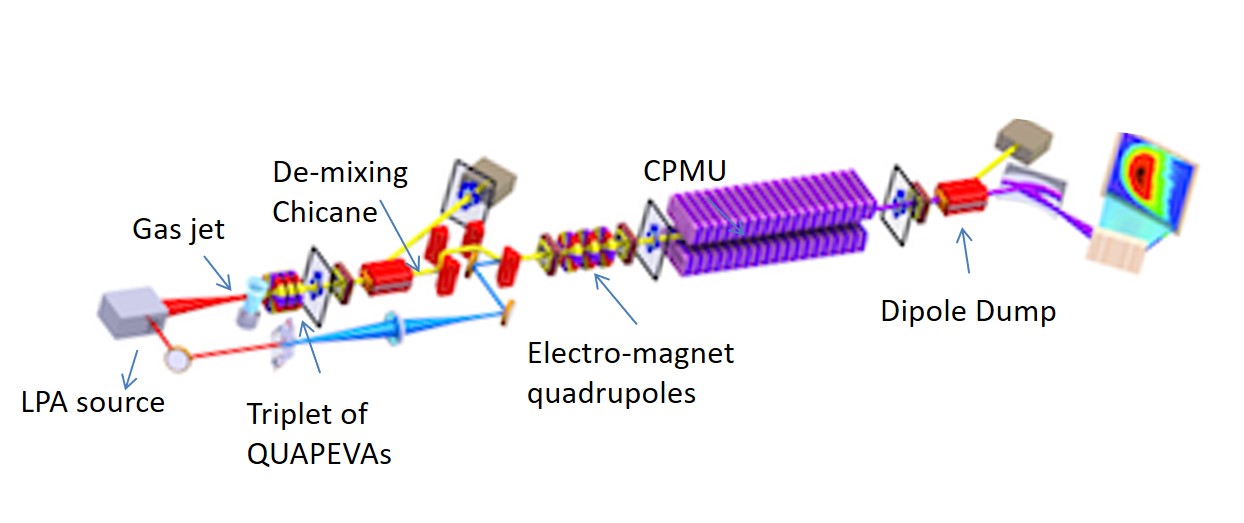}
\caption{\label{Fig7}COXINEL beam line magnetic elements.}
\end{figure}

A set of triplet of QUAPEVAs had been installed at COXINEL and enabled to control the beam up to 10 m long of transport line passing through different magnetic elements. Since the electron beam quality does not directly fulfill the FEL requirements, a transport line has been designed to manipulate the electron beam characteristics, as shown in Fig. \ref{Fig7}. The line starts with the triplet just after the gas jet where the electrons are produced, a de-mixing chicane composed of four electro-magnet dipoles to decompress the beam, followed by a set of four electro-magnet quadrupoles to ensure good focusing at the center of a undulator (2 m long with period length of 18 mm). Different diagnostics are installed along the transport line to measure the beam characteristics\cite{couprie2016application} at different locations.

\begin{figure}[ht]
\centering
\includegraphics[scale=0.05]{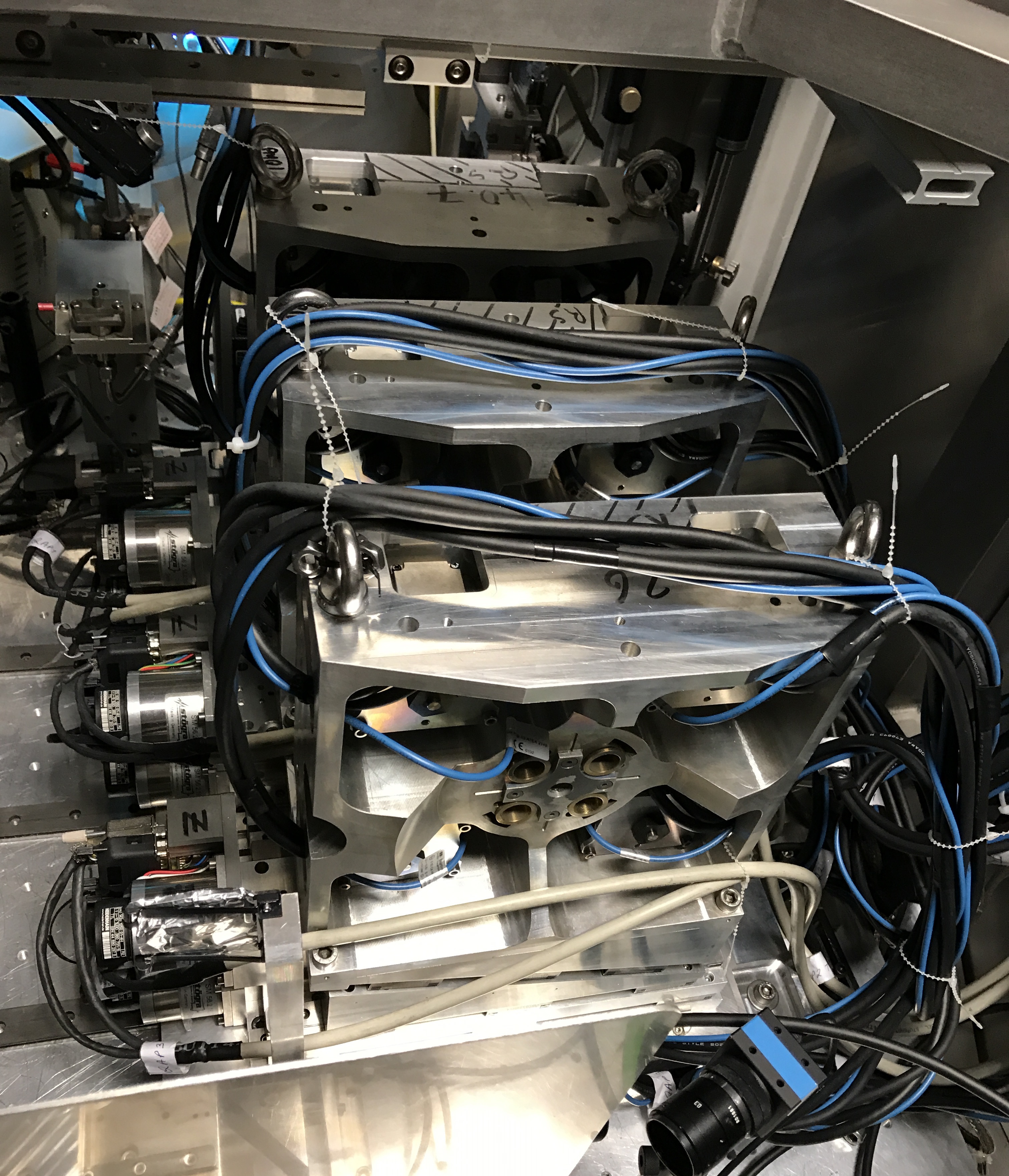}
\caption{QUAPEVAs triplet installed in COXINEL transport line.}
\label{Fig8}
\end{figure}

A view of the QUAPEVA installed in the gas jet chamber is shown in Fig. \ref{Fig8}. Since the QUAPEVA are very close to the focused laser, magnetic measurements have been performed after magnets use. In spite of their use in severe environments (laser and electron beam collision with the magnets), it was observed that there was no variation on the gradient or field quality.

\FloatBarrier

\begin{figure}[ht]
\centering
\includegraphics[scale=0.45]{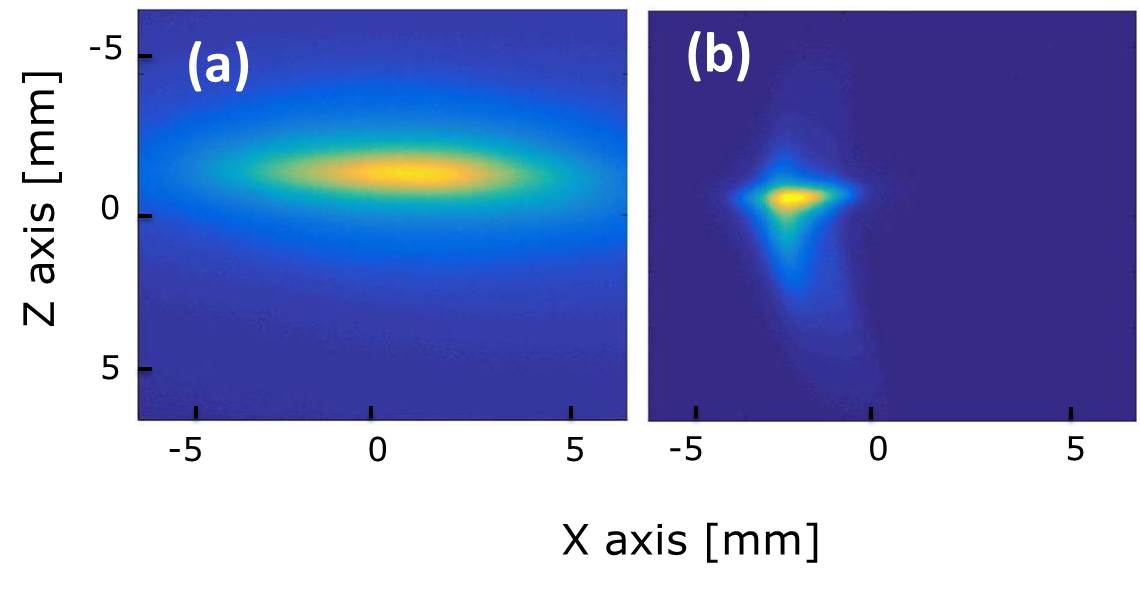}
\caption{(Left) the COXINEL scheme. (Right) electron beam (Energy: 176 MeV, divergence: 5 mrad) observation on first LANEX screen at 64 cm from the source : (a) Beam without QUAPEVAs, (b) Beam with QUAPEVAs}
\label{Fig9}
\end{figure}

 A beam observation on the first screen (LANEX) of COXINEL is shown in Fig. \ref{Fig9}, (a) without and (b) with QUAPEVAs installed. The large divergence of the electron beam is properly controlled and focused enabling beam transport further on the beam line.

\section{Conclusion}

The design of a permanent magnet based quadrupole of variable gradient strength have been presented. A high gradient ($\sim$ 201 T/m) with a wide tuning range ($\sim$ 92 T/m) is obtained with such a design combining a Halbach ring surrounded by four cylindrical magnets. The measurement using different methods are consistent and in good agreement. The residual excursion of the magnetic center has been limited to a $\pm 10 ~\mu m$ that can be compensated with the translation table. The quadrupoles have been installed successively at COXINEL beam line, and were able to achieve good focusing for a highly divergent large energy spread beam.

\section{Acknowledgments}

The authors are very grateful to the European Research Council for the advanced grant COXINEL (340015) and also to the Fondation de la Coop\'eration Scientifique for the Triangle de la Physique / valorisation contract QUAPEVA (2012-058T). The authors are very thankful for the COXINEL team as well, and acknowledge the Laboratoire d'Optique Appliqu\'ee team led by V. Malka for the generation of the electron beam by laser plasma acceleration, enabling to use the QUAPEVAs.

\end{document}